\documentclass{article}
\usepackage{spconf}
\usepackage{hyperref}
\usepackage{makecell}
\usepackage{graphics}
\usepackage{graphicx}
\usepackage{math}
\usepackage{xcolor}
\usepackage{url}
\usepackage[caption=false]{subfig}
\usepackage{amsmath}
\usepackage{amsfonts}
\usepackage{amssymb}
\usepackage[T1]{fontenc}
\title{Hearing in a Shoe-Box : Binaural Source Position and Wall Absorption\\  Estimation Using Virtually Supervised Learning\vspace{-2mm}}
\begin{document}
\urlstyle{same}
% Single address.
% ---------------
%\name{Author(s) Name(s)\thanks{Thanks to XYZ agency for funding.}}
\name{Saurabh Kataria$^{\star \dagger}$ \qquad Cl{\'e}ment Gaultier$^{\star}$ \qquad Antoine Deleforge$^{\star}$\vspace{-2mm}}

\address{$^{\star}$ Inria Rennes - Bretagne Atlantique, France \hspace{10mm}
$^{\dagger}$ Indian Institute of Technology Kanpur, India\vspace{-2mm}}
%
% For example:
% ------------
%\address{School\\
%   Department\\
%   Address}
%
% Two addresses (uncomment and modify for two-address case).
% ----------------------------------------------------------
%\twoauthors
%  {A. Author-one, B. Author-two\sthanks{Thanks to XYZ agency for funding.}}
%   {School A-B\\
%   Department A-B\\
%   Address A-B}
%  {C. Author-three, D. Author-four\sthanks{The fourth author performed the work
%   while at ...}}
%   {School C-D\\
%   Department C-D\\
%   Address C-D}
%
%\ninept
%
\maketitle
\begin{abstract}
\vspace{-1mm}
This paper introduces a new framework for supervised sound source localization referred to as virtually-supervised learning. An acoustic shoe-box room simulator is used to generate a large number of binaural single-source audio scenes. These scenes are used to build a dataset of spatial binaural features annotated with acoustic properties such as the 3D source position and the walls' absorption coefficients. A probabilistic high- to low-dimensional regression framework is used to learn a mapping from these features to the acoustic properties. Results indicate that this mapping successfully estimates the azimuth and elevation of new sources, but also their range and even the walls' absorption coefficients solely based on binaural signals. Results also reveal that incorporating random-diffusion effects in the data significantly improves the estimation of all parameters.
\end{abstract}


\vspace{-2mm}
\section{Introduction}
\vspace{-4mm}
Most existing methods in multichannel audio signal processing, including speech enhancement, denoising or source separation, rely on a good knowledge of the \textit{geometry} of the audio scene. In other words, what are the positions of the sources, sensors, and how does the sound propagate between them. Since this knowledge is most of the time unavailable, it is usually estimated from measured signals. A typical assumption is the free-field model in which the sound propagates in a straight line from each source to each sensor. Then, if sensor positions are known, sound source directions may be determined based on estimated time-differences of arrival. However, typical real-world audio scenes include reflecting and diffusive walls, floor, ceiling or filtering effects due, \textit{e.g.},  to the head of a binaural (2 microphones) receiver. Accurate audio scene geometry estimation is much more challenging in this context.

Two orthogonal research directions have recently emerged to tackle this challenge. The first one is \textit{physics-driven}, and consists in using more advanced acoustical models of the audio scenes. Such models may range from the image source model that incorporates specular reflections \cite{antonacci2012inference,dokmanic2013acoustic}, to the full wave propagation equation within boundaries of arbitrary shape and impedance \cite{bertin2016joint,kitic2016physics}. These methods are computationally intensive but yield encouraging results in simulated settings. The second direction is \textit{data-driven}, and consists in learning a mapping from measured high-dimensional acoustic features to source postions. Such mappings are learned from carefully recorded datasets in a supervised \cite{deleforge2013variational,deleforge2015co} or semi-supervised \cite{laufer2016semi} way. Since obtaining these datasets is time consuming, the methods are usually working well for one specific room and setup, and are hard to generalize in practice.

We now propose a third direction that somehow makes use of both worlds, namely, \textit{virtually supervised learning}. The idea is to use a physics-based room-acoustic simulator to generate arbitrary large datasets of audio-features in various geometrical settings. These data are then used to learn an efficient mapping from audio features to geometrical parameters. In this study, we make a first proof-of-concept by focusing on the scenario of a binaural receiver in a shoebox room of specific size. Over $80,000$ audio scenes with varying source direction, source distance, wall absorption, and random diffusion are generated using the ROOMSIM software of Shimmel and et al. \cite{schimmel2009fast}. We then extend the supervised sound-source localization method of \cite{deleforge2015co} to not only estimate 2D directions (azimuth, elevation) but also source ranges and mean wall absorption coefficients, solely based on binaural signals. Our experiments show promising results, and reveal that in the considered setting, the addition of diffusion effects significantly improve estimation of all parameters. While diffusion effects are most often neglected in the sound source localization literature, this suggests that they carry rich spatial information which may be helpful for binaural hearing.

\vspace{-2mm}
\section{Description of experimental setup}
\label{sec:2}
\vspace{-4mm}
The problem of single-source localization in a reverberant room using a binaural receiver is considered. It is well-known from both psychophysical \cite{hartmann1983localization,rakerd1985localization} and machine hearing \cite{deleforge2015co} studies that perceived binaural features do not only depend on the source's azimuth, but also on its elevation, its range, the position of the receiver in the room, and the room acoustic properties. The aim of this study is to investigate whether some of these additional parameters can be learned and estimated based on perceived spatial binaural features. The number and range of such parameters can be extremely vast considering the variety of real-world rooms, from anechoic chambers to cathedrals. Therefore, we choose to make a trade-off between realism and the number of parameters considered. The room size, receiver position and absorption profiles of the floor and ceiling are assumed fixed in all experiments. However, the azimuth, elevation and range of the source as well as the absorption profile of the walls (assumed identical for all walls), are varied. In addition, random-diffusion effects are added to account for the presence of sound scattering due to objects in the room. Finally, a white-noise emitter is used to avoid biases due to the specific spectral shapes of, \textit{e.g.}, speech signals. Details of the room simulation parameters are showed in Table~\ref{table:1}).

\begin{figure}[t!]
\resizebox{\columnwidth}{!}{
\includegraphics{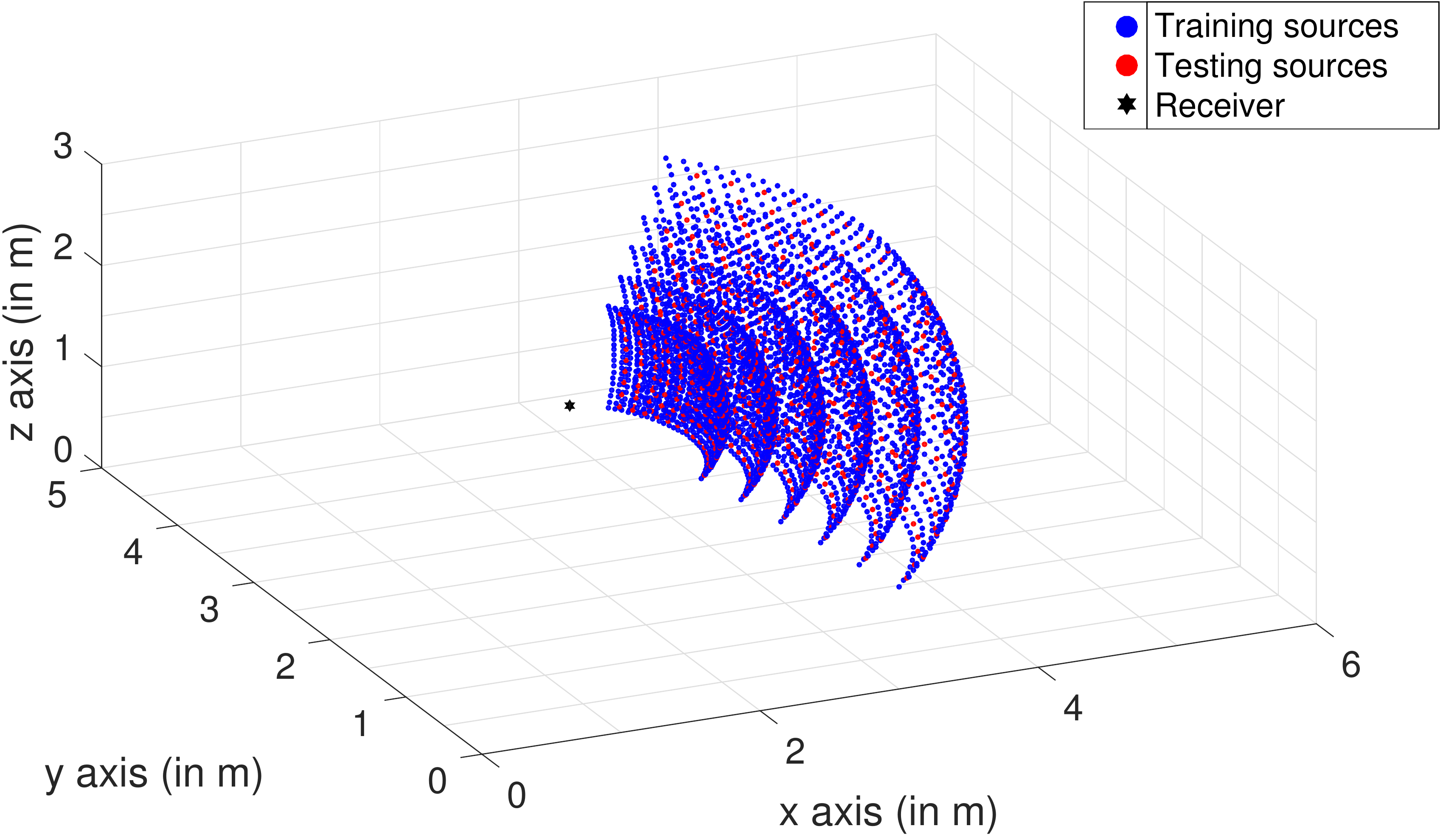}
}
\caption{\vspace{-8mm}\small Illustration of room setup \vspace{-8mm}}
\label{fig:1}
\vspace{-2mm}
\end{figure}

\begin{table}[t!]
\resizebox{\columnwidth}{!}{
\begin{tabular}{ |c|c| } 
 \hline
 Room dimensions (in m) & 6 x 5 x 3.3 \\
 \hline
 Receiver coordinates (in m) & (2, 2.5, 1.6)\\ 
 \hline
 Receiver HRTF model & MIT Kemar \cite{gardner1994hrft} \\
 \hline
 Sensor type & Omnidirectional \\
 \hline
 Six grid ranges (in m) & \{1, 1.3, 1.6, 1.9, 2.2, 2.5\} \\
 % \hline
 % Training sources per grid & 651 \\
 % \hline
 % Testing sources per grid & 150 \\
 \hline
 Angular range for sources (azimuth) & [-45$^{\circ}$, 45$^{\circ}$] \\
 \hline
 Angular range for sources (elevation) & [-30$^{\circ}$, 30$^{\circ}$] \\
 %\hline
 %\makecell{Duration of training \\ and testing signals} & 1.1 sec \\ 
 % \hline
 % Duration of testing signals & 1.1 sec \\
 \hline
 \makecell{Frequency bins from which \\ absorption and  diffusion profiles are \\ linearly interpolated (in KHz)} & \{0.125, 0.25, 0.5, 1, 2, 4\} \\
 \hline
 \makecell{Frequency-dependent absorption \\ values for ceiling (gypsum board)} & (0.45, 0.55, 0.60, 0.90, 0.86, 0.75) \\
 \hline
 \makecell{Frequency-dependent absorption \\ values for floor (thin carpet)} & (0.02, 0.04, 0.08, 0.20, 0.35, 0.40) \\
 \hline
 \makecell{Frequency-dependent diffusion \\ values for all surfaces} & (0.003, 0.004, 0.045, 0.077, 0.210, 0.431)\\
 \hline
\end{tabular}
}
\caption{\small General information about room setup.}
\label{table:1}
\vspace{-6mm}
\end{table}

The efficient C++/MATLAB ``shoebox'' 3D acoustic room simulator ROOMSIM developed by Schimmel et al. is selected for simulations\cite{schimmel2009fast}. This software takes as input a room dimension (width, depth and height), a source and receiver position, a receiver's head-related-transfer function (HRTF) model, and frequency-dependent absorption and diffusion coefficients for each surface. It outputs a corresponding pair of room impulse responses (RIR) at each ear of the binaural receiver. Specular reflections are modeled using the image-source method \cite{allen1979image}, while diffusion is modeled using the so-called \textit{rain-diffusion} algorithm. In the latter, sound rays uniformly sampled on the sphere are sent from the emitter and bounced on the walls according to specular laws, taking into account surface absorption. At each impact, each ray is also randomly bounced towards the receiver with a specified probability (the frequency-dependent \textit{diffusion coefficient} of the surface). The total received energy at each frequency is then aggregated using histograms. This model was notably showed to realistically account for sound scattering due to the presence of objects, by comparing simulated RIRs with measured ones \cite{wabnitz2010room}.

\begin{figure}[!t]

\centering
\subfloat[Absorption value of 18 materials with frequency\label{fig:mat_plots}]{%
  \includegraphics[scale = 0.28, keepaspectratio]{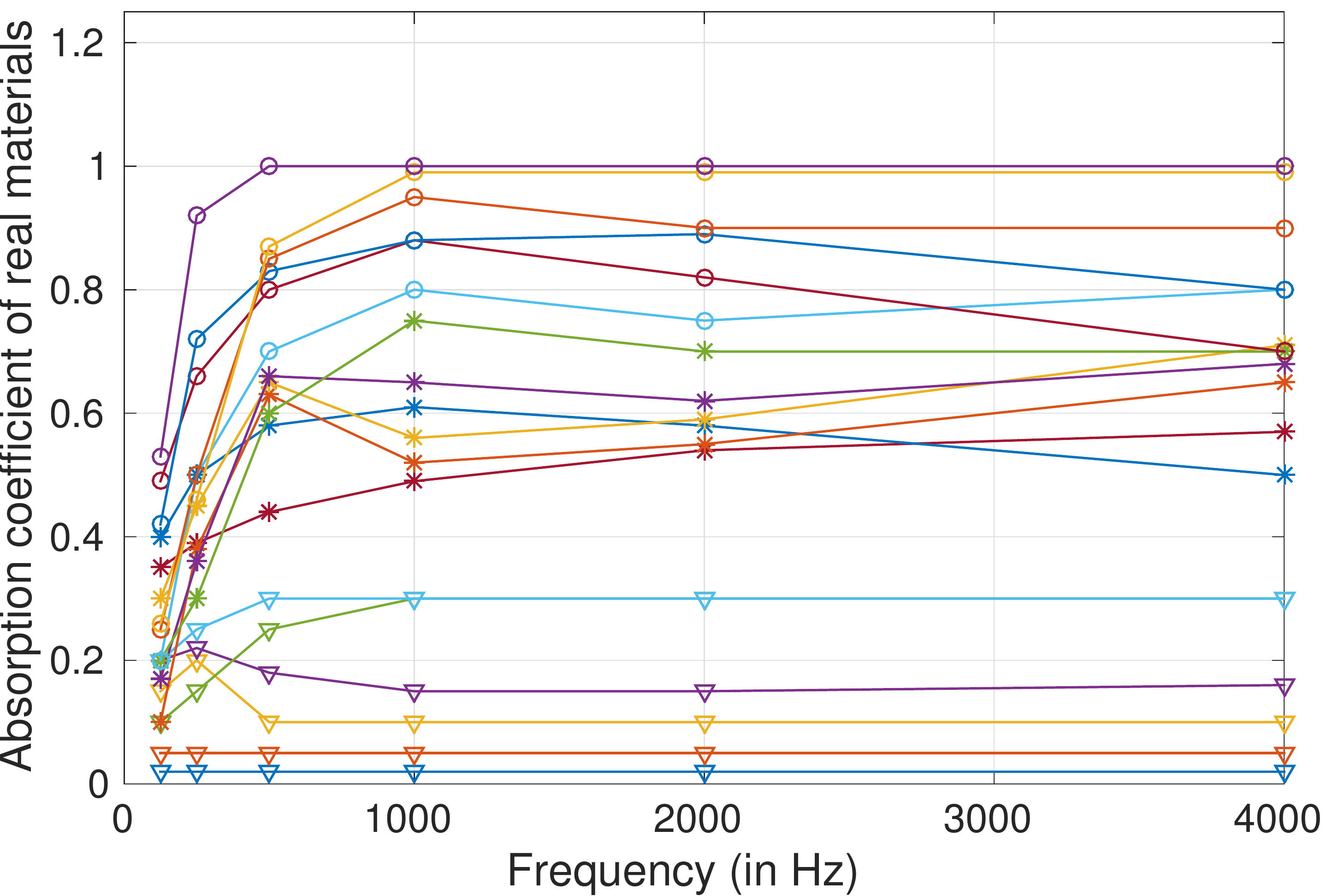}
}
\vspace{-2 mm}
\subfloat[Legend for the plot above\label{fig:mat_legend}]{%
  \includegraphics[scale = 0.28, keepaspectratio]{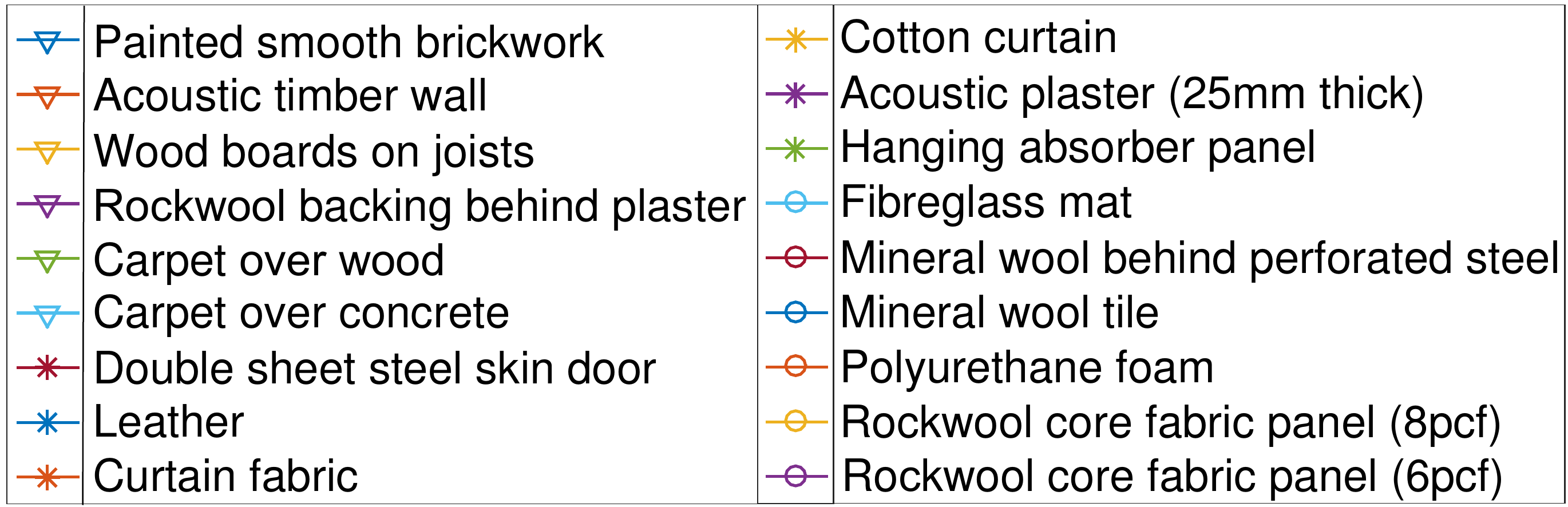}
}
\caption{\small Pictorial view of the dataset constructed}
\label{fig:materials}
\vspace{-6mm}
\end{figure}

The source positions considered are distributed on six spherical grids centered on the receiver, with radii vary from 1 meter to 2.5 meters. Each grid consists of 651 training positions and 150 distinct testing positions with uniformly distributed azimuths and elevations. The angular separation between consecutive positions is 3$^{\circ}$ for training and 6$^{\circ}$ for testing. Fig.~\ref{fig:1} shows the receiver and the different source positions within the room.

Since the absorption profile of a surface depends on frequency, the number of parameters required to fully model an absorption profile is too high, making its estimation unrealistic. A database of 18 measured absorption profiles of real materials was collected from \cite{vorlander2007auralization} and \cite{website2013}, and is listed in Fig.~\ref{fig:mat_legend}. As illustrated in Fig.~\ref{fig:mat_plots}, these materials show a strong absorption variability below 500 Hz. Then, the absorption slowly decreases or increases with frequency to reach a nearly constant value. These profiles were selected because the standard deviation (std) of their absorption is always below 0.07 between 500 Hz and 4 kHz. This allows to summarize the absorption profile of the four walls with a single parameter: the mean absorption coefficient above 500 Hz. Consequently, observations in frequencies under 500 Hz will be ignored in all experiments. Note that there is a second reason for ignoring these frequencies. The 60 dB reverberation times (RT60) measured in our experiments ranged from 0.09 to 1 second. This gives an upper-bound for the \textit{Schroeder frequency} of 200 Hz. The Schroeder frequency is given by $f_{\textrm{sch}}=2000\sqrt{\textrm{RT60}/V}$ where $V$ is the room volume in $m^3$, and  provides the frequency limit above which sound intensity is approximately homogeneous and isotropic \cite{bruneau2013fundamentals,kuttruff2009room}. Above that limit, the modal theory does not hold and is replaced by statistical models of diffuse fields. The simulator, which relies on such models, is thus more likely to give realistic results above that bound. The diffusion profile used in all experiments corresponds to the diffusion field added by three chairs, one table and one computer, as measured in \cite{faiz2012measurement}. As in, \textit{e.g.}, \cite{wabnitz2010room}, the same profile is used for all surfaces. Overall, each pair of generated RIR depends on four parameters: the source's azimuth, elevation and range, and the mean absorption coefficient of walls. The complete dataset is available online at \url{http://theVASTproject.inria.fr} (see also \cite{gaultier2016vast}).

\vspace{-2mm}
\section{Mapping Binaural Features to Geometrical Properties}
\vspace{-4mm}
\subsection{Computing Binaural features}
\vspace{-2mm}
Let $\uvect\in\mathbb{R}^4$ be a parameter vector containing the source's azimuth, elevation and range, and the mean wall absorption. We denote the associated generated left and right RIR by $(\hvect^\textrm{L}(\uvect),\hvect^\textrm{R}(\uvect))$. Each of these pairs is convolved with a 1 second random white Gaussian noise signal, and the result is resampled at 8kHz. The short-time Fourier transform is then applied to both signals, using a 64ms sliding time window with 50\% overlap. This results in a left-microphone spectrogram $\{L(f,t)\}_{f=1,t=1}^{F,T}$ and a right-microphone spectrogram $\{R(f,t)\}_{f=1,t=1}^{F,T}$, where $F=256$ and $T=32$. If $\{S(f,t)\}_{f=1,t=1}^{F,T}$ denotes the emitted white-noise spectrogram, under the assumption that most of the RIR energy is concentrated on the first 64ms, we have the following approximate multiplicative model
\begin{equation}
\label{eq:hrtf_model}
\left\{\begin{array}{l}
 L(f,t) \approx \hat{\hvect}^L(f,\uvect)S(f,t) \\
 R(f,t) \approx \hat{\hvect}^R(f,\uvect)S(f,t)
 \end{array}\right.
\end{equation}
where $\hat{\cdot}$ denotes the discrete Fourier transform. The \textit{interaural level difference} (ILD) and \textit{interaural phase difference} (IPD) spectrograms are defined by
\begin{equation}
\left\{\begin{array}{l}
 \operatorname{ILD}(f,t) = 20*\log(|L(f,t)|/|R(f,t)|) \in\mathbb{R} \\
 \operatorname{IPD}(f,t) =  \frac{L(f,t)/|L(f,t)|}{R(f,t)/|R(f,t)|} \in\mathbb{C}\equiv\mathbb{R}^2.
\end{array}\right.
\end{equation}
Using the approximation \eqref{eq:hrtf_model}, it is easily seen that both ILD and IPD solely depend on the parameter vector $\uvect$ and do not depend on the emitted signal. Similarly to \cite{deleforge2015co}, the ILD and IPD spectrograms are vertically concatenated and averaged over time to form a high-dimensional feature vector $\yvect\in\mathbb{R}^D$ associated to the low-dimensional parameter vector $\uvect\in\mathbb{R}^L (L=4)$. As explained in Section \ref{sec:2}, only the $F'=481$ bins corresponding to frequencies above 500 Hz are used in $\yvect$, resulting in a dimension $D=3F'=1443$ in practice.

\vspace{-2mm}
\subsection{Gaussian Locally-Linear Mapping}
\vspace{-2mm}
The training dataset is composed of $N$ pairs $\{(\yvect_n,\uvect_n)\}_{n=1}^N\subset\mathbb{R}^D\times\mathbb{R}^L$. A mapping needs to be learned from this dataset such that given a new test observation $\widetilde{\yvect}_t\in\mathbb{R}^D$, an associated parameter vector $\widetilde{\uvect}_t$ can be estimated. To achieve this, we use the high- to low-dimensional regression method \textit{Gaussian locally-linear mapping} (GLLiM) proposed in \cite{deleforge2015high}. GLLiM is a probabilistic method that estimates $K$ local affine transformation from the space of $\uvect$ to the space of $\yvect$ using a Gaussian mixture model. This mapping is then reversed through Bayes' inversion, yielding an efficient estimator of $\uvect$ given $\yvect$. GLLiM was successfully applied to supervised 2D sound source localization on a real dataset in \cite{deleforge2015co}. In practice, a fixed value $K=25$ is used in all experiments, as this showed to be a good trade-off between accuracy and computational time using preliminary validation sets.

\vspace{-2mm}
\section{Experimentation and results}
\label{sec:experi}
\vspace{-2mm}
We first conduct an experiment to reproduce the supervised binaural 2D localization results of \cite{deleforge2015co} in our setting. The GLLiM model is trained on 651 individual white noise (WN) recordings obtained from sources lying on a grid at a range of 1 m (the closest grid in Fig.~\ref{fig:1}). The chosen wall absorption profile is ``Rockwool backing behind plaster'' (Fig.~\ref{fig:mat_legend}) with mean absorption value 0.16. Random diffusion effects are added on both training and testing data. Two test cases are compared. First, testing in the same configuration, \textit{i.e.} , absorption and range are the same between training and testing. Second, testing with the different wall absorption profile ``Rockwool core fabric panel (8pcf)'' (Fig.~\ref{fig:mat_legend}, mean absorption value 0.96) and a grid range of 2.5 m. As expected, table \ref{table:2} illustrates that the localization error is higher when testing in a configuration different from the training configuration, in particular in elevation. The same phenomenon was observed in \cite{deleforge2015co} with real data. As in \cite{deleforge2015co}, it can be noted that estimating elevation is harder than azimuth, which is expected because of head symmetry. This difficulty seems further increased here by the use of a cutoff frequency at 500 Hz. This fundamental experiment motivates the idea of a robust training using multiple configurations of absorption values and grid ranges.

\vspace{-3mm}
\begin{table}[!htbp]
\small
\centering
\resizebox{\columnwidth}{!}{
\begin{tabular}{ |c|c|c|}
\hline
 & \makecell{Testing in the same \\ configuration} & \makecell{Testing in a different \\ configuration} \\
\hline
\makecell{Azimuthal error $({}^\circ)$} & $1.67 \pm 1.22$ & $1.99 \pm 1.42$ \\
\hline
\makecell{Elevation error $({}^\circ)$} & $8.78 \pm 7.08$ & $15.79 \pm 12.39$ \\
\hline
\end{tabular}
}
\caption{\small Comparing the mean and std of absolute localization errors when training with a single room and source range.}
\label{table:2}
\vspace{-2mm}
\end{table}

\begin{table*}[t!]
\small
\centering
% \resizebox{\columnwidth}{!}{
    \begin{tabular}{|c|c|c|c|c|c|}
        \hline
         \makecell{Training set\\annotation $\rightarrow$} & \makecell{Direction+Range+Ab-\\sorption} & \makecell{Direction+Range+Ab-\\sorption (no diffusion)} & Direction Only & Direction+Absorption & Direction+Range \\
         \hline
        Azimuth (${}^\circ$) & 1.78 $\pm$ 1.34 & 2.16 $\pm$ 1.62 & 1.72 $\pm$ 1.43 & 2.00 $\pm$ 1.51 & 1.91 $\pm$ 1.52\\
        \hline
        Elevation (${}^\circ$)  & 7.87 $\pm$ 6.45 & 11.3 $\pm$ 7.95  & 8.81 $\pm$ 7.81 & 8.45 $\pm$ 6.86 & 9.44 $\pm$ 7.55 \\
        \hline
        Range (cm)  & 54.2 $\pm$ 29.65 & 56.8 $\pm$ 34.3 & - & - & 58.5 $\pm$ 32.4 \\
        \hline        
        Absorption  & 0.18 $\pm$ 0.14 & 0.80 $\pm$ 0.44 & - & 0.22 $\pm$ 0.17 & -\\
        \hline
    \end{tabular}%
% }
\caption{\small Mean$\pm$std localization errors for various training sets. The second column is the same as first one but without diffusion. Vertical labels denote variables to be estimated. Horizontal labels denote parameters supplied during training (vector $\uvect$).}
\label{tab:allres}
\vspace{-4mm}
\end{table*}

We then trained the GLLiM model using all 651 training directions, 6 ranges (Fig.~\ref{fig:1}), and 21 constant wall absorption values in $\{0, 0.05,..., 1\}$. This results in a dataset of $N=651\times6\times21=82,026$ binaural feature vectors associated to $L=4$-dimensional parameter vectors $\{(\yvect_n,\uvect_n)\}_{n=1}^N\subset\mathbb{R}^D\times\mathbb{R}^L$. The 21 absorption values used here correspond to ideal materials with perfectly constant absorption coefficient in the frequency range $[0.5, 4]$ KHz.
Testing is done on the 108 test directions (Fig.~\ref{fig:1}), 6 grid ranges, and the 18 real material absorption profiles of Fig.~\ref{fig:mat_legend} and six grid ranges. By choosing ideal materials for training and real materials for testing, it is ensured that the training and testing sets are significantly different.

\begin{figure}[t]
\centering
\subfloat[\label{fig:comb-1}]{%
  \includegraphics[scale = 0.33, keepaspectratio]{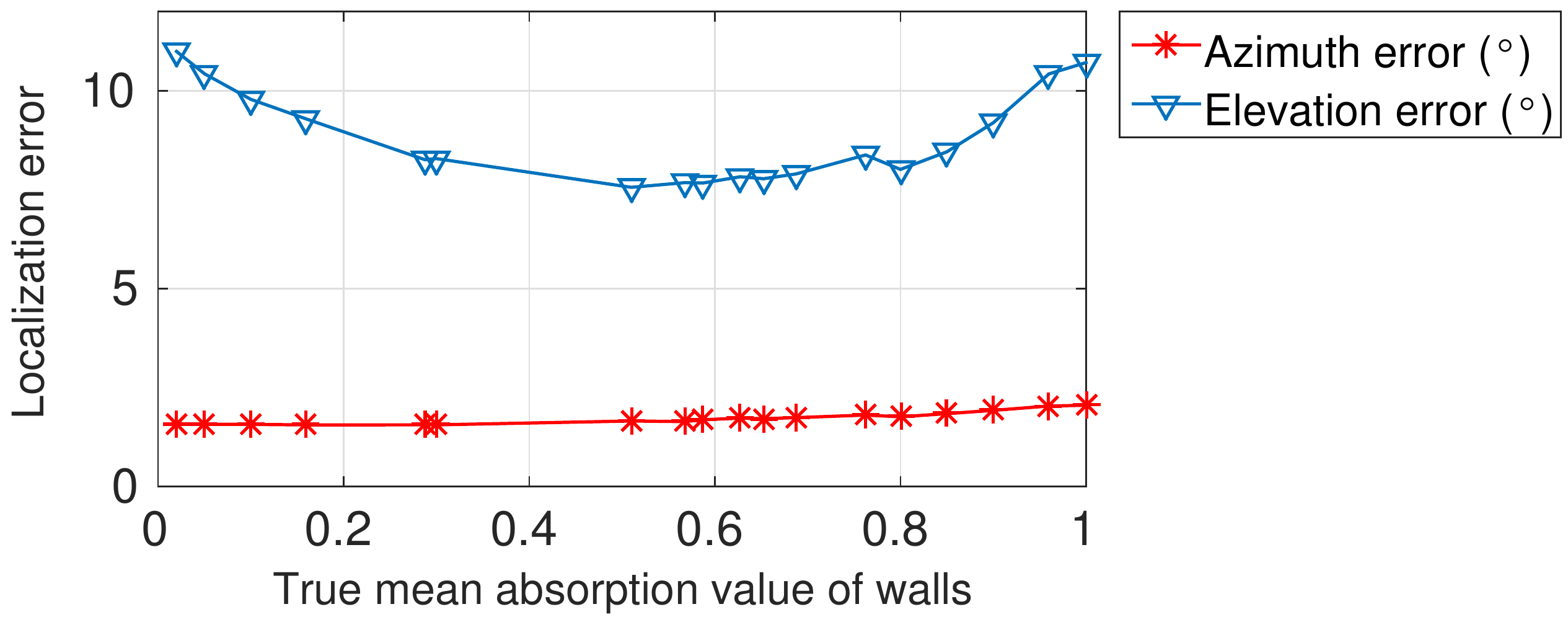}
}
\vspace{-4mm}
\subfloat[\label{fig:comb-2}]{%
  \includegraphics[scale = 0.32, keepaspectratio]{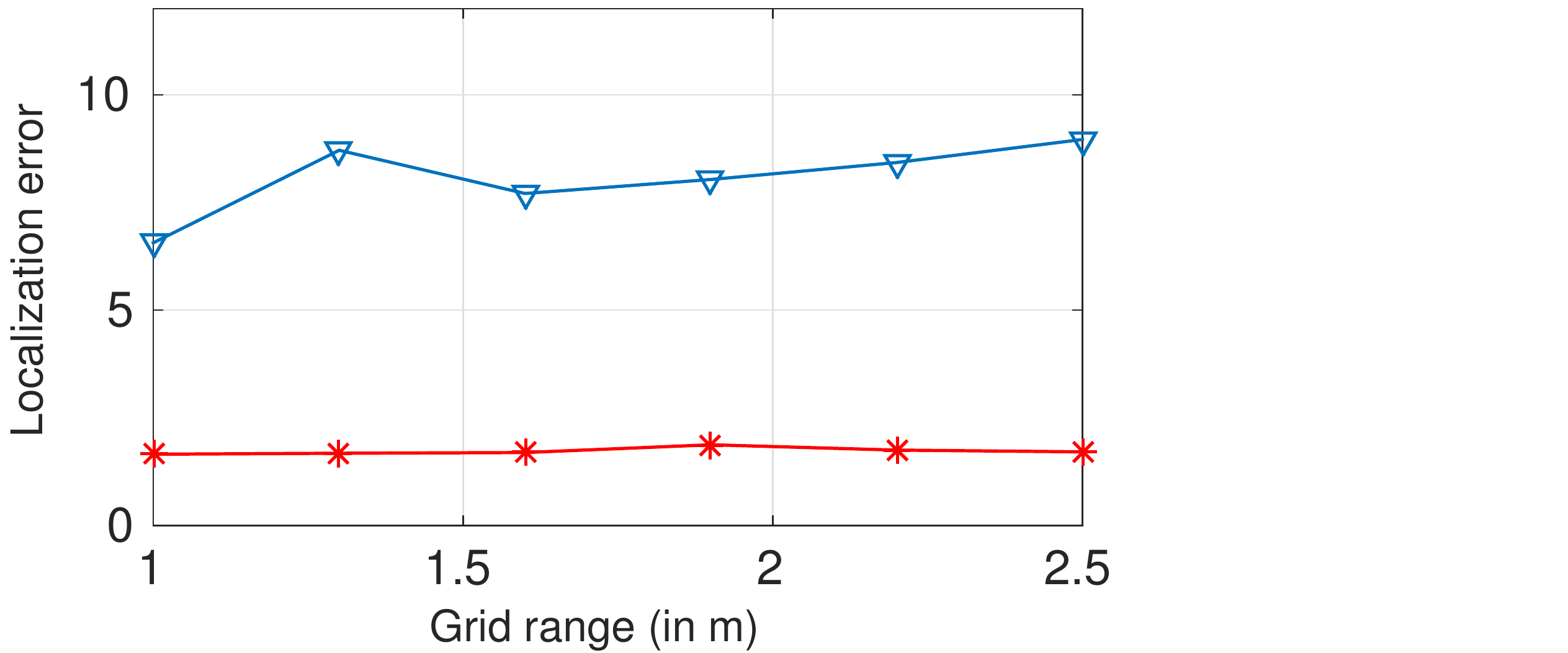}
}
\vspace{-2mm}
\caption{Mean localization error (in degrees) as a function of absorption (a) and range (b).}
% \label{fig:gridspec}
\vspace{-4mm}
\end{figure}

\vspace{-1mm}
Results are presented in the first column of Table~\ref{tab:allres}. As can be seen, training with the entire dataset improves localization results compared to training using a single room and range (Table \ref{table:2}). Moreover, our method is able to estimate the mean wall absorption between 0 and 1 with an accuracy of 0.18, as well as the range of the source between 1 and 2.5m with an accuracy of 54cm. To the best of the authors' knowledge, the only other binaural range estimation method in the literature is \cite{lu2010binaural}, which showed an accuracy of about 1m using direct-to-reverberant ratios. Figure \ref{fig:comb-1} shows mean source direction errors as a function of wall absorption. Note that too large or too little absorption hinders elevation estimation, the optimal results being obtained in the range $[0.3,0.8]$. This is probably because a high-level of reverberation implies that model \eqref{eq:hrtf_model} is more approximate, while too little reverberation means less spatial richness, and less symmetry breaking. Similarly, Fig.~\ref{fig:comb-2} shows errors as a function of source distance. As commonly observed, farther sources are slightly harder to localize, although the method seems to be particularly robust to range variations.

\vspace{-1mm}
Comparing now the first and second column of Table~\ref{tab:allres}, we observe an interesting result. When removing diffusion in both training and test data, the estimation of all variables is degraded. It was consistently observed that adding diffusion in simulations improved results, even when using a number of other diffusion profiles and different room dimensions. This is particularly striking for elevation and absorption. We suspect diffusion to increase the spectral richness of binaural cues, making them more discriminative by breaking inherent symmetries of the problem. The authors are not aware of previous work specifically studying this effect in the literature.

\vspace{-1mm}
We finally test the influence of removing absorption or range annotation during training (Columns 3 to 5 in Table~\ref{tab:allres}). As expected, removing this information increases errors, but overall, the GLLiM probabilistic framework seems to be relatively robust to additional non-annotated effects such as absorption and range. This is promising for future experiments on larger datasets with larger parameter variability, where full annotation of all effects may not necessarily be possible.

% \subsection{Absorption coefficient estimation}
% In this experiment, we are interested in estimating the mean absorption coefficient of walls of an unknown room in the frequency range [0.5, 4] KHz from a given test recording.
% 
% GLLiM is trained in universal-room universal-range type fashion. There is, however, one difference in training and testing procedure from the one mentioned in Section \ref{subsec:4.1}. Along with source location information, mean absorption coefficient value of walls is also provided. As a result, GLLiM is trained using a three-dimensional label vector $(\theta, \phi, a)$ and its corresponding binaural feature vector. Here, $\theta$ denotes the azimuth angle of source, $\phi$ denotes the elevation angle of source, and $a$ denotes the mean absorption coefficient of the walls. Results are reported in second column of Table \ref{tab:allres}.
% 
% \subsection{Source range estimation}
% In this experiment, we are interested in estimating range of source from a given test recording. GLLiM is trained in universal-room universal-range type fashion. Results for two cases are presented in Table \ref{tab:allres}. In first case, only range information is specified along with location. In second case, range as well as absorption value is specified. Results show that GLLiM is able to learn range estimation from binaural recordings in both diffusion and no diffusion environments.

\vspace{-4mm}
\section{Conclusion}
\vspace{-4mm}
In this paper, we presented a proof-of-concept for the novel framework of virtually-supervised learning for audio-scene geometry estimation. Obtained results are encouraging, revealing that estimating the 2D direction and range of a source as well as some of the wall acoustic properties is possible using binaural recordings only. Moreover, we observed that incorporating random diffusion effects in simulations significantly increased the spatial richness of binaural features, improving estimation of all parameters. In a simultaneous companion paper \cite{gaultier2016vast}, we validated the virtually-supervised learning framework by successfully localizing sources from \textbf{real-room signals} and showed the superiority of the approach over a traditional time-delay-based method. Extensions to speech \cite{deleforge2015co}, multiple sources \cite{deleforge2013variational} and additional partially-latent variables in GLLiM \cite{deleforge2015high} will also be considered.

\bibliographystyle{IEEEbib}
\small

\clearpage

\bibliography{refs}

\end{document}